# Self-assembled Frameworks Solid with Turbostratic Stacked Crystalline Layers - A Frustrated 3D Crystal Lattice

Hongmei Qin,[1,] Jiahui Wang,[1] Na Lin, [1] Xiaoxu Sun, Yin Chen [1,*]

College Chemistry & Chemical Engineering, Central South University, Changsha 410083, Hunan, China;

*Correspondence to: chenyin@iccas.ac.cn.

**Abstract**

 Solid materials possess both long-range order and some degree of disorder are critical for understanding the nature of crystal and glassy state, but how to controllably introduce specific type of disorder into a crystalline material is a big challenge. Our previous work indicated that weaking the interlayer interaction is an effective strategy to import disorders between the layers. Here, we illustrated that the interlayer interaction can be weakened to around 1/60 of that of graphite in the self-assembled IPM-31, a 2D frameworks formed by bicyclocalix[2]arene[2]triazines tricarboxylic acid with $Cu^{2+}$ nodes, which has an obvious layered-structure. It's difficult to obtain IPM-31 in 3D crystalline form, but an amorphous state turbostratic solid with irregular appearance, electron microscopy images clearly show that IPM-31 was piled up by lots of nanosheets with high flatness. However, the in-plane long-term periodicity in turbostratic IPM-31 has been ambiguously detected by electron microscope, electron diffraction and Synchrotron Radiation XRD etc. The theoretical study identified that the extremely weak interlayer interaction is the key factor for the formation of turbostratic structure, which leads to the frustration of regular 3D lattice. This interlayer interaction can be further weakened by small ligand exchange, as a result, spherical particle form IPM-31 was produced, which is an aggerate of IPM-31 nanosheets with completely random and twisted arrangement. For the same reason, IPM-31 can be can be easily exfoliated into ultra-thin nanosheets with high homogeneity and large lateral size.

For nearly 2000 years, the external morphology is predominantly concerned for crystals. Until 17th century, modern crystallography was established on the idea that crystal shapes were the result of internal order of "atomic" units. Based on the establishment of X-Ray diffraction crystallography, scientists were confident about that until the 1980s: A crystal is a regular repeating arrangement of unit cells. Different to the crystal, amorphous materials usually don't have both regular appearance and intrinsic three-dimensional long-range periodicity.[1-2] However, only very limited is known about the structure of amorphous phase materials,[3-4] due to the difficult experimental challenge not only in characterizing the disorder but also in the interpretation of data.[5-7]

To understand the nature of amorphous material, and clarify the boundary of crystal and amorphous state, it's critical to find or design some material structure with both long-rang order and some degree of disorders.[8] The nature and phase behavior of disordered crystalline states have long been studied within the statistical mechanical and crystallographic communities, there are many known partially-ordered materials. Prof. B. E. Warren is one of the pioneers in the research of amorphous materials with X-ray diffraction.[9] In the study of carbon black, he realized that it is not completely amorphous since there is two-dimensional periodicity within the graphene layer, but it also does not have the three-dimensional periodicity of graphite. In hence, carbon black was supposed to be an intermediate form of matter, and the term "turbostratic" (unordered layers) is suggested as a name for this particular mesomorphic solid, which possess both long-range order and disorder.[10] However, carbon black usually was produced uncontrollably under extreme conditions, it is still very challenging to produce the turbostratic carbon controllably.[11-12]

On the other hand, the relationship between the disorder and property of solid material is an emerging and challenging frontier topic.[13] There are several key issues involved, and one of the main challenges is how to controllably introduce a specific type of disorder into a crystalline material. In the next, proper and advanced characterization methods are badly-needed to understand the form of that disorder, as well as whether it can be introduced as-designed. At last, it is very important to figure out the possible physical or chemical properties origins from the disorders. Previous reports exhibited there are several different ways to introduce the disorders into a crystal. At first, disorders can be produced by employing different components or defect to occupy the same crystallographic site, such as the TRUMOF-1,[14] the prussian blue analogues with general formula as $M^{II}[M^{III}(CN)_6]_{2/3} \square _{1/3}$ ($\square$ = vacancy), $AuAg(CN)_2$,[15] hcp UiO-67.[16] Alternatively, the disorder also generated via directed self-assembly, such as the spontaneous defects observed in the p-Terphenyl-3,5,3′,5′-tetracarboxylic acid molecules assemble,[17] columns assemble of 1,3,5-benzenetrisamides.[18]

In many other cases, disorders were caused by the thermal fluctuations, or the existence of defects and grain boundaries. Materials that fall in this category include superconducting cuprates and colossal magneto resistive manganates, in which there is substitutional disorder, and grossly nonstoichiometric binary and ternary compounds.[19] However, the conventional X-rays or neutron diffraction often fails to reveal the precise nature of the disorder. Systems with apparent long-range order can also contain defects resulting from domain formation, it is still a very challenging question at which point the structure ceases to be crystalline.

Metal-organic frameworks (MOFs) are framework materials of great current interest, in particular because of their unique and tailorable porosities.[20-21] Generally, MOFs are consisted of nodes of metal atoms or clusters linked together by organic molecular ligands, which endow great freedom for their structure design,[22] disorder can be controllably introduced in turn.[14, 23] Most of current MOFs are synthesized in the form of three-dimensional (3D) crystals. Recently, we have found that bicyclocalix[2]arene[2]triazines, a cage-like building block, tends self-assemble into layered 2D metal-organic frameworks with metal cation nodes.[24-25] What's more interesting is that, the cage-like building block leads to the extremely low interlayer real contact area between the stereo porous layers, only very weak interaction can be found between the neighbour layers.[24-29] The interaction can be further adjusted by varying the metal atom nodes and the cage ligands. Due to the molecular thermal motion, relative interlayer sliding took place when the interlayer interaction is weak enough, leading to directed disorder between the layers.[24, 30-31] In hence, it can be concluded that weakening the interlayer interaction is an effective strategy to introduce disorder into the crystalline lattice in specific dimension.[32] Then, there is an interesting question that what will happen with even weaker interlayer interaction?

Here, we found that bicyclocalix[2]arene[2]triazines tri-carboxylic acid (**BCTA**) can self-assemble with $Cu^{2+}$ into a layered 2D frameworks, IPM-31, which has extremely weak interlayer interaction. Based on the single crystal structure, the interlayer interaction was estimated to be 1/60 of that of graphite based on DFT calculation. Experimental results show that it's very difficult to obtain IPM-31 with regular crystal shape, and conventional in-house XRD even found this material as amorphous.[33] However, the electron microscopy images clearly show that IPM-31 solid is stacked by ultra-thin nanosheets with irregular appearance, which fits very well with the definition of turbostratic structure. The intrinsic long-term periodicity of "amorphous" IPM-31 can be detected by the synchrotron radiation XRD, as well as the transmission electron microscope (TEM) and electron diffraction, which confirmed the ultra-thin nanosheets have consistent periodicity as the single crystal structure. Due to the critical weak interlayer interaction, it's difficult for IPM-31 to form stable 3D lattice, but a frustrated turbostratic structure, which possess both long-range order and significant disorders between neighbour layers. Unsurprisingly, IPM-31 can be easily exfoliated into ultra-thin nanosheets with high thickness homogeneity and large lateral size.[34-35]

**IPM-31** was prepared from **BCTA** and Cu(NO₃)₂ in anhydrous dimethyl sulfoxide (DMSO) under 120

ºC with a yield around 70%. Under general reaction conditions, the light blue IPM-31 solid show an

amorphous powder XRD pattern, and not suitable for single crystal XRD (SCXRD). Under vibration-free

and precise temperature controll condition, IPM-31 crystal suitable for SCXRD can be obtained. IPM-31

crystallizes in the triclinic space group P -1 (Table S1, Supporting Information). In the asymmetric unit,

there are one **BCTA** ligand and one Cu(II) atom. Two neighboring Cu(II) atoms are bridged by four

carboxylate groups from four different but symmetrically equivalent BCTA to form a bi-paddle-wheel

[Cu₂(O₂C−)₄] SBU.

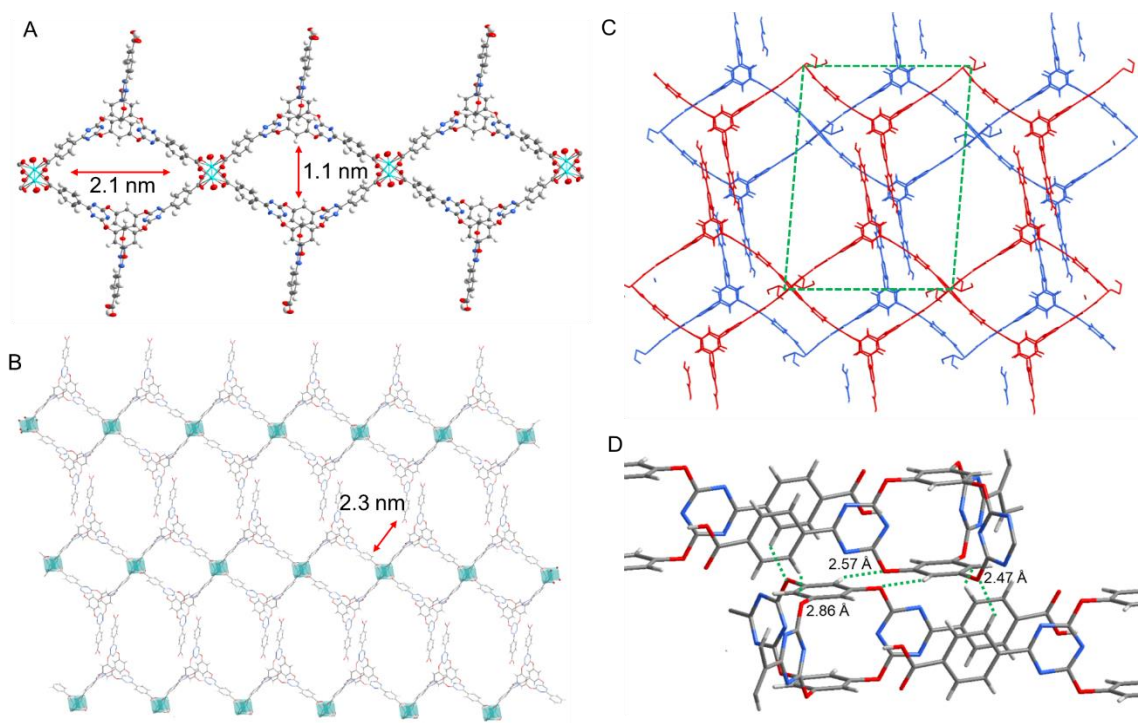

**Figure 1**. The layered structure of the IPM-31 single crystal. (A) The 1D chain structure formed by BCTA

and the bi-paddle-wheel [Cu₂(O₂C−)₄] SBU with rhombus cavities (2.1 nm * 1.1 nm); (B) The 2D layer

assembled by the 1D chain via π···π stacking between carboxylic acid arms, there are hexagon cavities with

edge length around 2.3 nm between two neighbor chains; (C) The stacking mode between two adjacent 2D

layers (red and blue), with very low real contact area between the layers (The green parallelogram shows the basic interaction unit with an area of 6.8 nm$^2$); (D) The C-H$\cdots\pi$ and C-H$\cdots$O interaction sites and bonding mode observed in one basic unit.

The two copper nodes of the SBU further coordinates with two dimethyl sulfoxide molecules to form octahedron configuration. **BCTA** has three equivalent carboxylic acid ligand arms, but only two carboxyl groups are deprotonated. Each BCTA molecule links two bi-paddle-wheel SBUs to form a 1D chain structure (Figure 1A) with tandem rhombus cavities (2.1 nm * 1.1 nm). The spare arms on the chain interact with the other two chains in the neighborhood mainly via multiple $\pi\cdots\pi$ stacking in a mode similar to the DNA structure, leading to a 2D layer with a thickness around 0.65 nm (Figure 1B, S1). The periodical interaction sites significantly enhanced the stability of the self-assembly structure, and hexagon shaped cavities can be found between two parallel chains.

The net-like layers staggered stacked parallel to the crystallographic *c*-axis to form the 3D crystal lattice (Figure 1C, S2), due to the large-porous network, the real contact area between two adjacent steric layers is extremely low (Figure S3). Between the layers, only weak interaction between phloroglucinol benzene rings and arm benzene rings can be found. In an average 6.8 nm$^2$ area, two C-H$\cdots\pi$ interaction sites ($d_{C-H\cdots\pi}$ =2.86 Å) and four C-H$\cdots$O interaction sites (two sets, $d_{C-H\cdots\pi}$ =2.47, 2.57 Å) can be observed (Figure 1D). The dimethyl sulfoxide coordinated with the [Cu$_2$(O$_2$C$-$)$_4$] SBU wedged into the cavities of adjacent layers, which contribute to the stabilization of crystal lattice. As shown in the photomicrograph, although IPM-31 has a transparent appearance close to the ordinary crystals (Figure 2A, S4), the in-house powder XRD always record a typical amorphous diffraction pattern for IPM-31 prepared in large batch (Figure 2B). However, scanning electron microscopy (SEM) images confirmed that **IPM-31** has the consistent layered structure as observed in single crystal structure, but this solid was piled up by lots of distinct and smooth nanosheets (Fig. 2C, 2D, S5), which arranged randomly in translation parallel to the layer, and also rotated about the normal. The layered structure of IPM-31 also can be in the transmission electron microscope (TEM) images (Figure 3A, S6). Element mapping (Figure 3B) images revealed a homogenous distribution of C, N, O and Cu elements throughout the IPM-31 samples, indicating a consistence of composition in this solid. The intrinsic periodicity of the IPM-31 nanosheets can be directly detected by the high-resolution (HR) TEM, lattice fringes with the spacing range around 0.53 nm can be clearly found (Figure 3C, S7), which can match the simulated result based on the single crystal. Selected area electron diffraction (SAED) also confirmed the multiple twinned crystal nature of IPM-31, as typical polycrystal diffraction pattern was observed. These experimental results unambiguously exhibited the periodicity of IPM-31, which is the turbostratic aggregate of crystalline nanosheets instead of common amorphous material based on the classic X-ray diffraction crystallography.

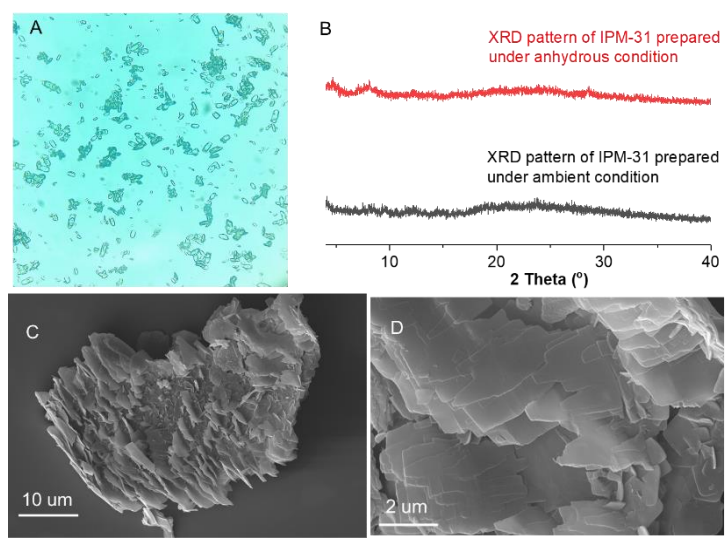

**Figure 2**. (A) The photomicrograph of IPM-31 prepared in large batch; (B) The typical amorphous powder diffraction pattern recorded for IPM-31; (C) The SEM image of IPM-31 show its turbostratic structure; (D) The nanosheets aggregate texture of IPM-31 observed in the amplified SEM image.

The periodicity of IPM-31 was further characterized by synchrotron-based in situ X-ray diffraction. The experiment was conducted by the BL14B1 beamline at Shanghai Synchrotron Radiation Facility (SSRF). Corresponding X-ray beam possesses the wavelength of 0.6887 Å with an energy of 18 KeV. The XRD patterns were collected by a Mython 1K detector with a distance of 330 mm between samples and detector. As shown in Figure 4A, diffraction peaks were observed at $2\theta = 2.3^o$, $3.6^o$, $3.9^o$ and $6.1^o$, which can be attributed to the reflection of (010), (10$\bar{1}$), (1$\bar{1}$0) and (120) planes, respectively, Figure S8-11). (010), (10$\bar{1}$), (1$\bar{1}$0) planes all are periodical element with the IPM-31 layers [Cu$_2$(O$_2$C$-$)$_4$] SBU, (120) plane is the central plane of the IPM-31 layer. It can be concluded that all these diffraction peaks can be the 2D periodicity in the IPM-31 layers and the layer , which confirmed the 2D periodical structure of IPM-31.

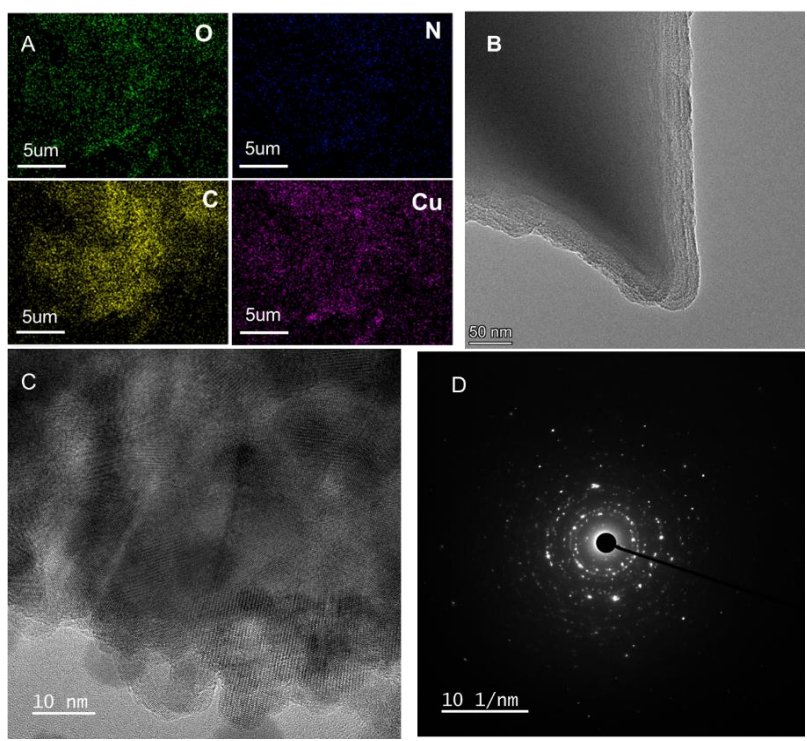

**Figure 3**. (A) Element mapping image of IPM-31 showing homogeneous distributions of O (green), N (blue), C (yellow) and Cu (purple) elements; (B) The TEM image show the layered texture of IPM-31; (C) Lattice fringes with different orientations observed in the HR-TEM image of IPM-31, the spacing ranges is around 0.54 nm ; (D) The SAED patten of IPM-31 show its polycrystal nature.

Thermal gravity (TG) analysis data show that IPM-31 is thermally stable up to 280 $^{\circ}$C. A weight loss around 17% under 120 $^{\circ}$C can be attributed to the removal of free solvent molecules in the cavities (Figure S12).  IPM-31 has a typical type V $N_2$ adsorption-desorption isotherm, indicating a very weak interaction with the gas molecules based on the capillary condensation theory. The calculated BET surface area for IPM-31 is only 68 m$^2$/g (after activation at 120 $^{\circ}$C under vacuum, as shown in Figure S13). IPM-31 shows a very broad pore-size distribution (Figure S14) without detectable micropores. One reasonable explanation for these contradictory experimental results is that, the random arrangement of the layers has destroyed and blocked the regular channels observed in the single crystal structure of IPM-31, resulting to the ambiguous pore-size distribution.

Due to the extremely low interlayer interaction, **IPM-31** is readily to be exfoliated into monolayer nanosheets *via* routine ultrasonic exfoliation. A milk white **IPM-31** nanosheet suspension can be easily produced in large batch, significant Tyndall effect was observed (Figure S15). In this suspension, lots of ultra-thin and smooth nanosheet with lateral size range from several to more than ten micrometers can be observed in the TEM images (Figure 4B, S16). Tapping-mode atomic force microscopy (AFM) further confirmed this point, Figure 4C shows an enlarged area of the IPM-31 nanosheet, the corresponding height profile reveals that the nanosheet is extremely flat, with an even step height around 1.58 nm, it can be concluded that the IPM-31 nanosheet is around 2-3 layers thick. As demonstrated by the AFM analyses on 36 different sites (Figure 4D, S17), most of these nanosheets have a thickness of 2.1±0.6 nm with a high thickness homogeneity.

Another interesting fact is that, after expose to moisture, the color of IPM-31 will change from light blue to green after some time. According to previous reports, it's well known that the bi-paddle-wheel [$Cu_2(O_2C-)_4$] SBU has a high chemical and structural stability,[36-38] which remains stable even in the boiling water. Our previous research also illustrated the even higher stability of BCTA. In this case, the possible cause for the color change is the ligand exchange of the copper sites. As mentioned above, the coordinated DMSO molecules have played an important role in stabilizing the IPM-31 lattice, it's logical to suppose that the replacement of DMSO by the water molecule will further weaken the interlayer interaction. Therefore, the preparation of IPM-31 was performed with undried DMSO, which failed to obtain the turbostratic IPM-31 as expected. Instead, green-colored irregular spherical particles was produced with a size up to some hundreds of micrometers (Figure 5A, S18), which also give a typical amorphous diffraction pattern (Figure 2B, black). However, the SEM images found that these spherical particles also were piled by ultra-thin nanosheets, but in an even random and distorted manner (Figure 5B, 5C, S19).

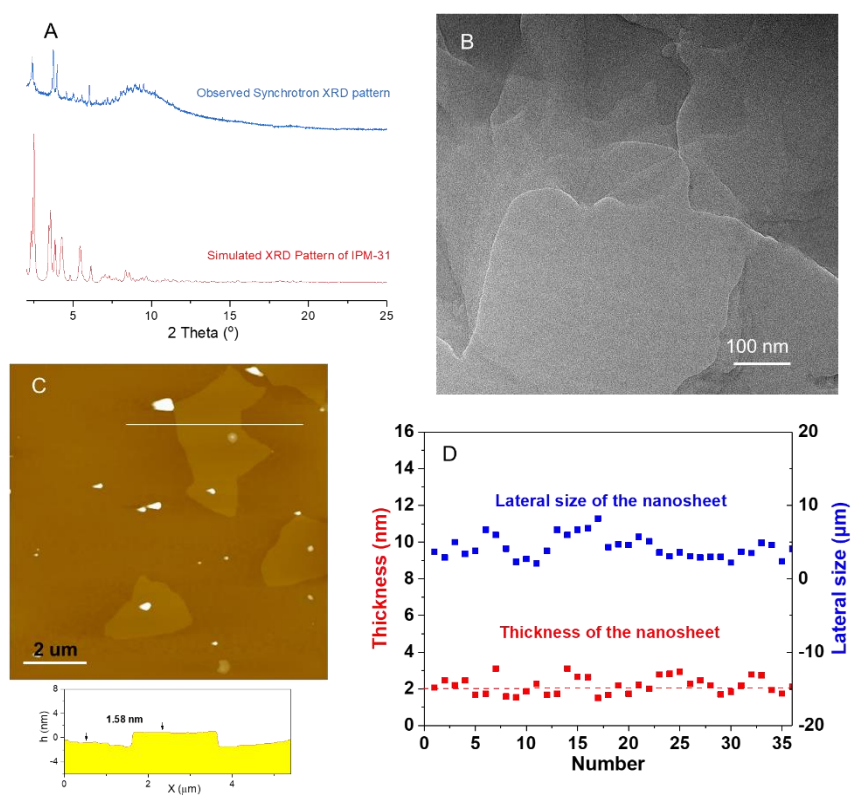

**Figure 4**. A) Synchrotron radiation X-ray diffraction pattern of IPM-31. B) TEM image of **IPM-31** nanosheets. (C) Tapping-mode AFM image of **IPM-31** nanosheet, the height profile of the nanosheets was recorded along the white lines marked in the AFM image. (D) Thickness and lateral size distribution of the exfoliated **IPM-31** nanosheets (horizontal black dotted line indicates the theoretical thickness of a single layered **IPM-31** nanosheets).

The synchrotron radiation XRD confirmed that the spherical particle has the same poor diffraction pattern as the turbostratic IPM-31 after expose to moisture, the original diffraction peaks of IPM-31 remained but with much weaker strength, many other very weak diffraction peaks also can be found, indicating the existence of periodicity in the 2D layers. But the extremely weakened interlayer interaction leads to the failure of self-assembly of turbostratic structure.

After coordination with $Cu^{2+}$ to form IPM-31, Fourier transform infrared (FT-IR) spectra show the characterization absorbance bands of **BCTA** at 1687 $cm^{-1}$, 1531 $cm^{-1}$, 1286 $cm^{-1}$, 877 $cm^{-1}$ and 780 $cm^{-1}$ have disappeared, and the absorbance bands at 1565 $cm^{-1}$ and 1164 $cm^{-1}$ have significantly changed in strength (Figure S20). Turbostratic and spherical IPM-31 have very similar IR spectra, indicating an identical chemical composition. Turbostratic IPM-31 has an extra band around 1090 $cm^{-1}$, which can be attributed to the IR absorbance of DMSO. It was further supported that the coordinated DMSO caused the morphology difference of IPM-31.

To further understand the experimental results, density functional theory (DFT) calculation was employed to estimate the specific interaction intensity between the IPM-31 layers, as well as the influence of the coordinated DMSO. Considering the crystallization process of IPM-31, obviously, the driving force for the formation of 2D IPM-31 layers is stronger that the 3D stacking. It's reasonably to suppose that the formation of the 2D layer crystal nucleus is kinetically advantaged. The formation of crystal or turbostratic structure is decided by the combination energy between two IPM-31 layers, which is defined by the formula: $E_{com} = E_{total} - 2E_{IPM-31}$. $E_{total}$ is the total energy of two combined IPM-31 layers, $E_{IPM-31}$ is the individual energies of one single IPM-31 layer in DMSO. The negative adsorption energies indicate the attractive noncovalent interactions, which are common in the adsorption of molecules on the surface of materials.

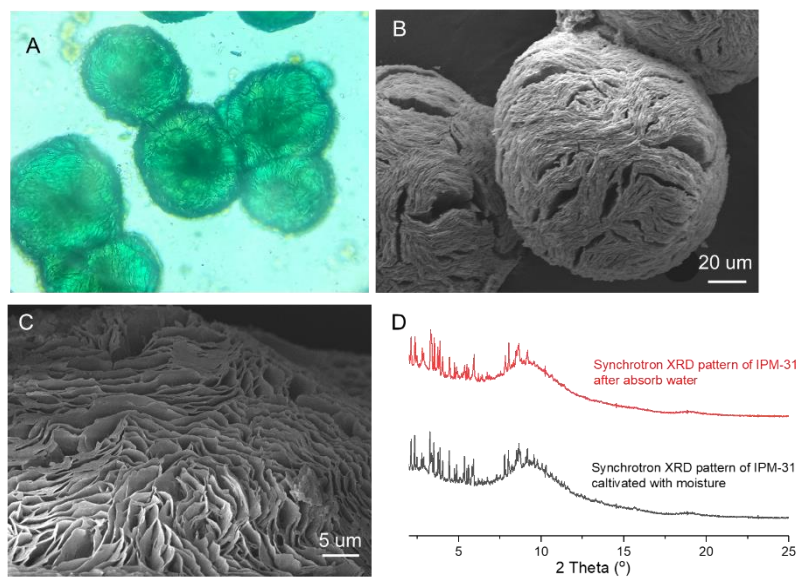

**Figure 5**. (A) The green colored spherical IPM-31 produced in the undried DMSO; (B), (C) The SEM image of spherical particle IPM-31 identified its nanosheet aggregate structure; D). Synchrotron radiation X-ray diffraction pattern of IPM-31 after absorb water and spherical IPM-31.

The extent of charge transfer, *i.e.*, the difference in charge between pristine IPM-31 layers and the IPM-31 layer-DMSO complex, was estimated by Bader charge analysis. This value reflects the amount of change in the valence electron of IPM-31, which also is negative. Compare to the IPM-31 layer-DMSO complex, the estimated $E_{com}$ is 21.8 kJ/mole (approximately $5.3 \times 10^{-21}$ J for 1 nm$^2$ contact area), which is less than 1/60 of the binding energy between the graphene layers. Indicating a very small energy barriers to overcome the lattice energy, which provides a good theoretical explanation for the formation of turbostratic structure, the interlayer interaction is too weak to hold the 3D lattice structure, leading to the frustrated 3D crystal lattice.

In order to understand the influence of coordinated DMSO to the stability of the lattice, the structure of IPM-31 layers with coordinated DMSO or $H_2O$ were optimized, the resulting $E_{com}$ values and charge transfer per unit area were calculated as shown in figure S17. The replacement of DMSO by $H_2O$ caused a 12 kJ/mole decrease of the interlayer sliding energy barrier. As a result, the 2D layer crystal nucleus prefers growing randomly in the solvent instead of stacking on each other.

In summary, weakening the interaction along specific dimension in the crystal lattice is an effective strategy to introduce directed disorder, here, we have investigated an extreme situation when the interlayer interaction is too weak to hold adjacent layers, which leads to the frustration of 3D crystal lattice. To be specific, IPM-31, a 2D frameworks self-assembled from **BCTA** and $Cu^{2+}$ nodes, has an interlayer interaction estimated to be $5.3 \times 10^{-21}$ J per 1 $nm^2$, which is only around 1/60 of that of graphite. As a result, it's difficult to obtain IPM-31 in the common crystal form, but a turbostratic structure solid with irregular appearance. IPM-31 should be classified as amorphous material according to current definition. Meanwhile, the SEM, TEM, AFM, SEAD and Synchrotron Radiation XRD characterizations ambiguously identified the existence of in-plane long-term periodicity in IPM-31. "Turbostratic" is a term generally used in carbon black material, but the mechanism and the driving force for producing such material is still far from well-understood. To our best knowledge, IPM-31 is the first material in this kind can be prepared in a controllable approach. The theoretical study confirmed that the critical weak interlayer interaction is the key factor for the formation of turbostratic structure, as it is too weak to form stable 3D lattice. In another words, turbostratic structure is a frustrated 3D crystal. Unsurprisingly, IPM-31 can be easily exfoliated into ultra-thin nanosheets with high thickness homogeneity and large lateral size, suggesting that the design of 2D frameworks with extremely low interlayer interaction is a promising strategy for the synthesis of new low-dimensional materials. In the other hand, it also can be concluded that the frameworks look amorphous by in-house PXRD may not always be attributed to an experimental failure.

**Acknowledgments:** The authors acknowledge the resources and facilities provided by College of Chemistry &

Chemical Engineering of Central South University. Y. C. thanks the help in Synchrotron Radiation XRD provided

by Prof. Wen Wen and Prof. Su Zhenhuang from SSRF. Y. C. thanks the financial support from Central South

University, State Grid Shaanxi Electric Power Research Institute. The authors thank beamline BL14B1 at the

Shanghai Synchrotron Radiation Facility (SSRF) for providing the beam time.

**Competing interests:** The authors declare no competing financial interests.